\documentclass[a4paper]{jpconf}
\usepackage{graphicx}
\begin{document}
\title{Dynamical processes in galaxy centers}

\author{Francoise Combes}

\address{Observatoire de Paris, LERMA, CNRS, 61 Av de l'Observatoire, 75014, Paris}

\ead{francoise.combes@obspm.fr}

\begin{abstract}
How does the gas get in nuclear regions to fuel black holes?
How efficient is the feedback? 
The different processes to cause rapid gas inflow (or outflow) in
galaxy centers are reviewed. Non axisymmetries can be created
or maintained by internal disk instabilities, or galaxy interactions. 
Simulations and observations tell us that the fueling is a chaotic
and intermittent process, with different scenarios and time-scales,
according to the various radial scales across a galaxy.
\end{abstract}

\section{Introduction}

Black hole growth in the center of galaxies appears to
be closely related to the growth of bulges, either through
star formation or mergers (Magorrian et al 1998, G\"ultekin et al 2009).
Analysis of the BH demography today shows that $\sim$ 25\% of
the black holes mass is contained in late-type galaxies; the
most massive BH have grown faster, at higher redshifts, and
the BH mass growth through mergers is not dominant (Shankar et al 2004). 
 Most of the BH growth occurs in type 2 AGN, obscured by dust,
which supports the association with star formation.

The dynamical processes able to produce non-axisymmetries 
in galaxy disks and consequently gravity torques on the gas,
driving inflows towards the center, are multiple, and 
depend strongly on the scale considered. At kiloparsecs scale,
stellar bars are the most robust and efficient instabilities,
they can be re-inforced or triggered by galaxy interactions. 
Below one kiloparsec, their action can be continued 
through secondary, nuclear bars, decoupling and interacting
non-linearly with the primary bars.
 At 10-100pc scale, nuclear disks may become unstable
through $m=1$ modes, lopsidedness, warps and bending.
Below that, the clumpiness of the interstellar medium
plays a large role so that dynamical friction, and also
viscous torques due to turbulence could take over.

In recent years, it has been realised that
feedback phenomena, due to star formation or the AGN itself, 
can produce substantial gas outflows, self-regulating
the inflow and the fueling. In the following, all
these processes are discussed.

\section{Bar gravity torques}

When a stellar bar has grown in a galaxy disk,
it triggers spiral arms in the gas, which tends
to follow periodic orbits but dissipates energy at
crossing flows. The gas response and the stellar
bar are not in phase, which results in
gravity torques exerted on the gas, negative
inside corotation, and positive outside
(Garcia-Burillo et al. 2005, 2009).
 The torques are efficient to drive the gas from corotation to the 
center, or down to a ring at inner Lindblad resonance (ILR), which
is then the site of a starburst.

The amplitude of the torques and their efficiency 
can be computed from the observations:
the gravitational potential is derived from the
near-infrared image, and the tangential forces 
are computed on the gas distribution, allowing
to draw the torque maps, sometimes
showing negative torques down to the center (e.g. Casasola et al 2008). 
According to the bar strength, the gas can lose 10-30\% of its angular
momentum in one rotation, i.e. the inflow time-scale
is of the order of 300 Myr.
Note that the stellar potential can also exert
torques on the stars themselves, but this is much weaker, with time-scales of 4 Gyr
(e.g. Foyle et al 2010).

During the gas inflow inside corotation, the 
gas transfers its angular momentum to the stars
(the gas exerts a reciprocal torque on the bar),
and this weakens or destroys the bar, which is 
a wave with negative angular momentum
(Bournaud \& Combes 2002).
  When external gas accretion is taken into account, 
a new bar instability occurs in a disk replenished
in gas, and a cycle alternating bar formation and AGN 
fueling is triggered (Bournaud et al 2005).

 The fact that torques change sign at each resonance
prevents the gas to inflow in only one step,
all across the disk. Clearly,
several dynamical processes have to contribute
at the various scales.  Inside corotation, there are
frequently ILRs, and the gas piles up in rings
about these radii. 
When the mass concentration is sufficient in the
center, the frequency of orbital precession
($\Omega-\kappa/2$) is large enough that a nuclear
bar can decouple from the primary one. 
There is interaction between the two bars through resonances, 
for instance the corotation of the nuclear bar
is the ILR of the primary bar.
The secondary bar can then continue the
gas inflow towards the very center. 
This double inflow has been seen in
the type 2 AGN NGC2782, and has been reproduced
in fitted simulations (Hunt et al 2008).
It can be shown that the actual AGN fueling
is a very short phase, i.e. when torques are negative
down to the very center. If the gas continues to inflow
inside 200pc during almost 1 Gyr, it inflows inside 20pc only
during 100 Myr.

\section{Instabilities m=1}

Lopsidedness may occur at large scale in the outer
parts of galaxies, but also in the very center
(see e.g. Jog \& Combes 2009). A central mass concentration
favors the latter, since orbits tend to be keplerian.
In a thin self-gravitating disk, the $m=1$ resonant
orbits have a 
very rapid differential precession rate $\Omega-\kappa$,
while in the presence of a massive black hole, the potential 
becomes spherical and $\Omega$ = $\kappa$.
In these nearly keplerian disks, slow $m=1$ modes can develop,
an example being the conspicuous stellar disk around the 
black hole in M31. This mode develops
with a strong self-gravity, and the action of the
indirect potential, due to the off-centring of the central BH mass
(Bacon et al 2001, Tremaine 2001).

When gas is present in the disk, $m=1$ asymmetry torques allow the gas
to lose angular momentum, and flow to the center. This lopsidedness
is frequent at small scales in nearby galaxies, where spatial resolution 
is  sufficient (our Milky Way is an example), and larger samples have been 
studied at large scales. Over thousands of galaxies, a correlation has been
found between the lopsidedness, BH growth and the youth of stellar populations
in the center (Reichard et al 2009).

Recent simulations by Hopkins et al (2011) illustrate all these
dynamical processes, that occur at decreasing scales. After 
a cascade of $m=2$ and $m=1$ perturbations, the gas is focussing
towards the center, and forms a thick disk of size 1-10pc around
the black hole.
Zoomed simulations allow to go further, and study
the fate of the gas that has piled up in the nucleus,
with a gas fraction up to 90\%. 
Column densities can then reach 10$^{22}$-10$^{25}$ cm$^{-2}$,
the gas is unstable to formation of clumps, but also
warps, twists and bending.
At this stage, dynamical friction of Giant Molecular Clouds (GMC)
of M= 10$^6$ M$_\odot$ against the bulge can occur in a time-scale
t$\sim$ 80Myr $(r/100pc)^2$, at a distance $r$ from the center,
and varies in 1/M.

\begin{figure}[ht]
\begin{centering}
\includegraphics[width=15cm]{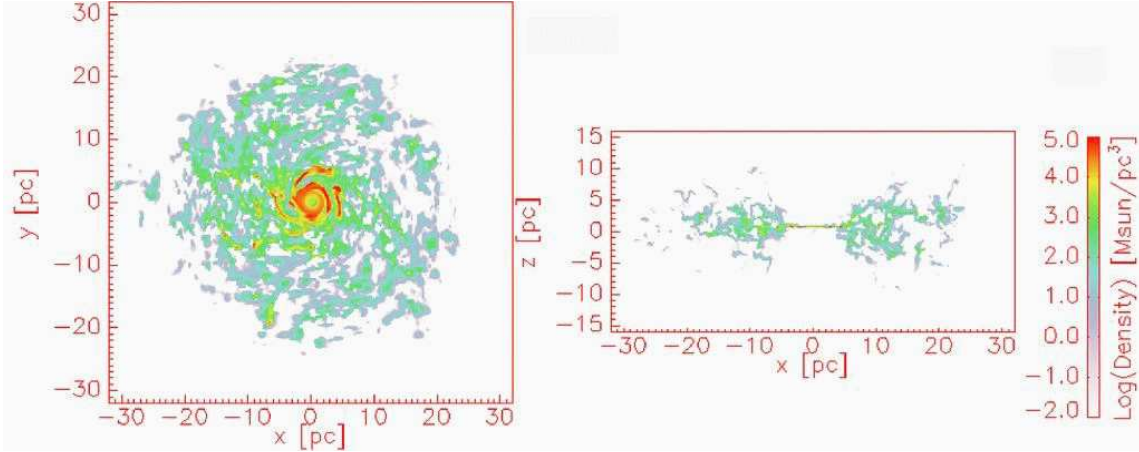}
\caption{\label{fig1}
 The molecular gas distribution in the galaxy center, in the X-Y plane (left)
and in a perpendicular slice (X-Z), in a torus simulation including H$_2$ formation
by Wada et al (2009).}
\end{centering}
\end{figure}

\section{Formation of the torus}

The resulting dense, compact and thick disk around the BH could be
identified as the molecular torus, often invoked to explain type 2 AGN.
The warp and bending instabilities can be maintained
through continuous gas accretion, and this explains why
the torus and outer disk in galaxies are frequently misaligned.
The morphology and thickness of the torus obtained depends strongly
on the physics assumed, and in particular on the velocity 
dispersion, the heating or the star formation feedback adopted, in general
at a scale below the resolution of the simulations.

Both warm thin and cold thick, fluffly, disks/torus could co-exist, according
to the multi-phase simulations by Schartmann et al (2009).
Under the influence of stellar winds, from
a nuclear stellar cluster, a nuclear disk of radius 10pc
forms with cold filaments and a tenuous hot interstellar medium.

Thinner molecular disks are obtained by Wada et al (2009),
through introducing the formation of molecular hydrogen (H$_2$),
and consistent chemistry, in presence of an UV flux of
G=10 to 100 in Habing units. They simulate a disk of 30pc radius,
heated by supernovae feedback, with a  finest resolution of 0.125pc.
The H$_2$ fraction at equilibrium is 0.4. The disk is
extremely thin in the central 5pc, but then flares at a radius of 10pc,
as shown in Figure 1.

The influence of stellar feedback is of course crucial on
the morphology of the disks/torus. 
Hopkins \& Quataert (2010) have introduced an equation of state
of the gas, to take into account the subgrid physics, and in particular
the consequence of star formation on the dipersion and turbulence of the gas.
Exploring the whole SN efficiency, the simulations can reproduce
the physical properties (density, dispersion) observed for
various objects, from quiescent disks to perturbed ULIRGs.

They show through several successive re-simulations
how the inflow rate 
has a chaotic evolution at all scales (cf Figure 2).
The accretion is episodic, and is much more contrasted in a merger,
than during secular evolution.

\begin{figure}[ht]
\begin{centering}
\includegraphics[width=15cm]{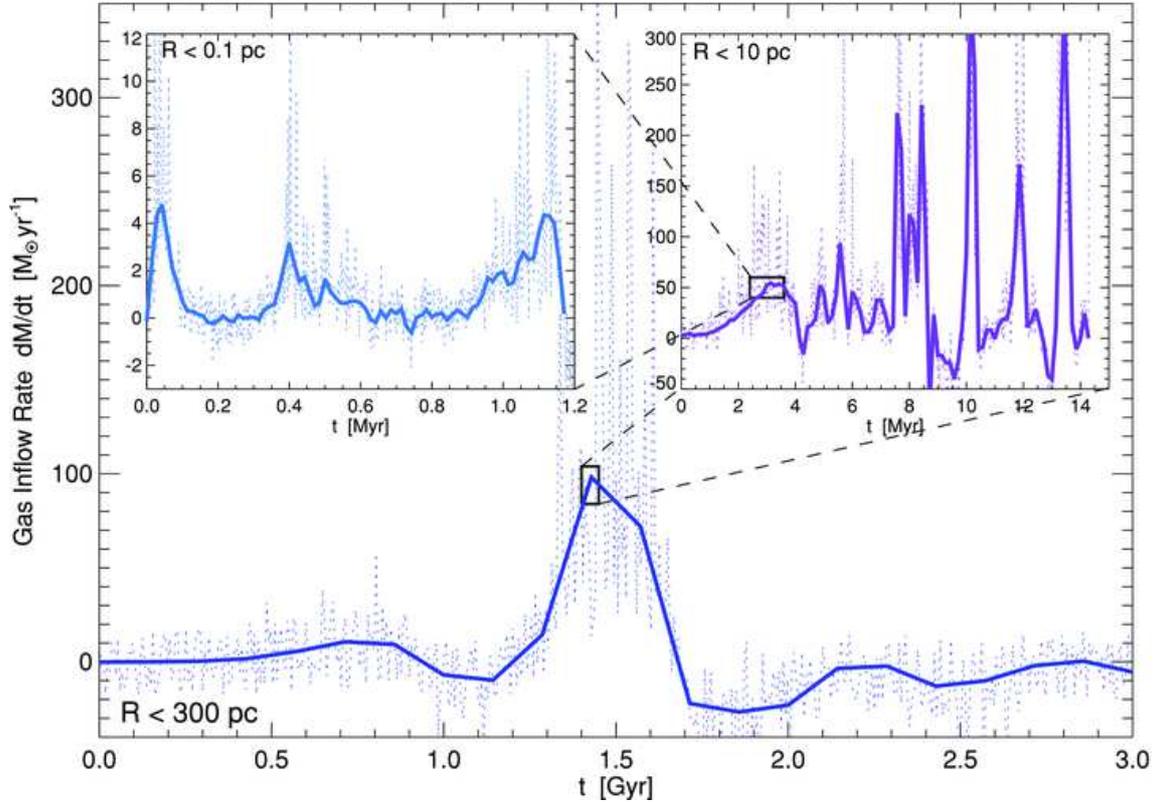}
\caption{\label{fig2}
 The inflow rate as a function of time, during the merger of two galaxies,
and in particular the coalescence of the two nuclei. 
The main simulation (bottom) shows the accretion inside 300pc,  and
is resimulated twice at higher spatial and time 
resolution (top right inflow inside 10 pc
and top left inside 0.1 pc). The dash line is the instantaneous rate, 
smoothed to 5 local dynamical times in the full line, 
from  Hopkins \& Quataert (2010).}
\end{centering}
\end{figure}

\section{Quasar feedback}

In the standard model of galaxy formation, it is quite
difficult to reproduce the galaxy mass function, both at the
small mass and large mass ends of the spectrum. For dwarf galaxies,
supernovae feedback has long been invoked to eject baryons out
of the shallow potential wells, but this does not work for massive galaxies.
For the latter, AGN feedback seems appropriate, given the
BH-bulge mass correlation, although no global
observation has yet confirmed its efficiency.
Simulations (e.g. Di Matteo et al 2005) have shown how in galaxy mergers
the energy released by the AGN quenches
both star formation and AGN growth.
Evidence of regulating or quenching star formation is observed
in special cases of central galaxies in clusters,
where radio jets moderate cooling flows.

Recently, more evidence has been obtained of
gas outflows in AGN/starburst galaxies. Many examples
of OH and H$_2$O line absorptions have been detected with
the Herschel satellite, a prototype being 
Mrk 231 (Fischer et al 2010).
In the same object, wide wings ($\sim$ 800 km/s)
in the CO emission have been detected with the 
IRAM interferometer (Feruglio et al 2010). The ejected
gas amounts to 700 M$_\odot$/yr, more than 3 times 
the star formation rate of 200 M$_\odot$/yr.
These neutral flows correspond to massive ejections,
while previoulsy many more ionized gas flows had been detected,
but with less mass involved. For example, the 
 [Fe II] emission line in Mrk 1157 has been observed with a kinematic
corresponding to outflow along the radio jet, which liberates
the iron from dust grains in shocks. The ejected ionized gas 
amounts to 6 M$_\odot$/yr (Riffel \& Storchi-Bergmann 2011).

Selecting optical AGN observed in the H$_2$ rotational lines with Spitzer,
we have recently found several cases of broad wings, implying molecular gas
outflows. In 4C12.50 for example (cf Figure 3), there is an H$_2$ component
flowing at 640km/s, and corresponding to an outflow
rate of 130 M$_\odot$/yr, much lower than the star formation rate, however
(Dasyra \& Combes 2011).

\begin{figure}[ht]
\begin{centering}
\includegraphics[width=10cm]{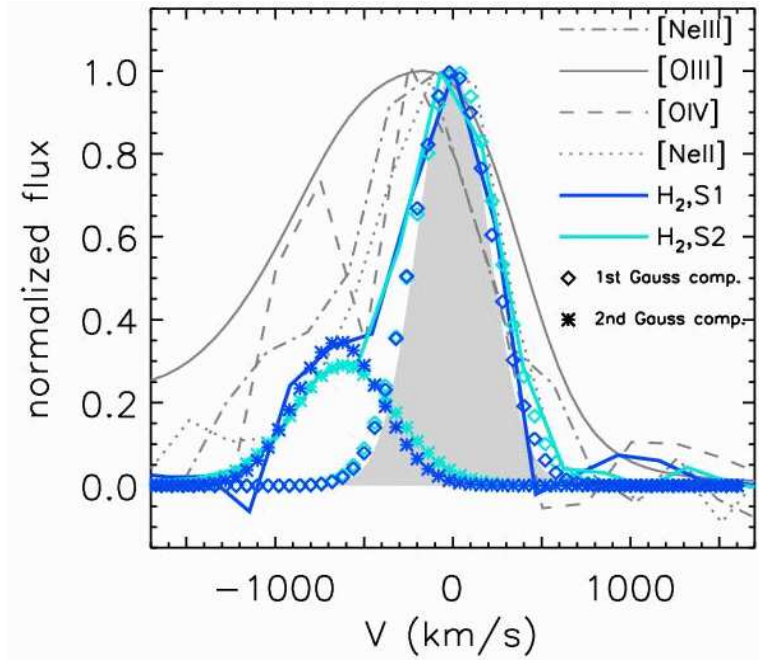}
\caption{\label{fig3}
Profiles of several lines, and in particular H$_2$ S1 and S2, towards
the quasar/starburst  4C12.50, showing broad line wings,
implying massive gas outflow, from Dasyra \& Combes (2011).}
\end{centering}
\end{figure}

Feedback might be invoked also to reduce AGN activity to a 
rapid phase in galaxy mergers. Indeed, given the frequency of galaxy mergers,
the presence of a massive BH in each galaxy center, and the gas inflow
in galaxy interactions, binary AGN are predicted to be more frequent than 
observed. Recently Green et al (2010) have studied a rare example of
binary quasar in a merger at z=0.44, and suggest that many others could be
obscured by dust.
Locally, binary AGN are very rare: NGC6240 has been 
shown to host two active nuclei through obscured X-ray emission and 
two strong neutral Fe K$\alpha$ lines (Komossa et al 2003).
The same evidence has been found for Arp 299 (Ballo et al 2004).

\section{Statistics - time-scales, 10-100pc fueling}

Among a sample of 16 nearby active galaxies with low-luminosity
(Seyfert or Liners), which was observed at high resolution at multi-wavelengths,
and in the molecular component with the IRAM interferometer, it is possible
to derive preliminary statistics about their fueling state. From their
torque maps, the current gas (in/out)flows have been derived 
(see also Garcia-Burillo, this meeting). 
Only $\sim$35\% (or 6 out of 16 galaxies of the NUGA sample) have 
 negative torques in the center, observed at a resolution
scale of 1"$\sim$50-100pc.
Most of the remaining galaxies show
positive torques towards the center, maintaining the gas in a nuclear ring.

This must imply that fueling phases are short,
a few 10$^7$ yrs, possibly due to feedback. This order of magnitude
is compatible to what is found in N-body simulations through bar fueling,
as presented above, and also to the rarity of binary AGN.
 With so short fueling phases, it is 
difficult to identify the trigger, bars or nuclear bars might have weaken
when the gas finally reaches the center. 
This explains the difficulty to find correlations between AGN
activity and a given dynamical process.
Star formation is also fueled by the same torques, 
is frequently associated to AGN activity, and is characterised by
longer time-scales, therefore correlations are easier to detect with
their dynamical trigger.

\section{Conclusion}

Dynamical processes depend essentially on the radial scale 
in galaxies.
Primary bars are essential to drive gas from 10kpc to R $\sim$ 100pc.
Then nuclear bars take over from 100pc to 10pc.
These bars could be part of the internal secular evolution, 
maintained by external gas accretion, or could be triggered 
by galaxy interactions.
Non-axisymmetries of the $m=1$ type, off-centring and lopsidedness,
may play a role, in the outer parts (gas accretion and tides), or
in the very center, when the potential is nearly keplerian, 
close to the black hole. In most cases, there are embedded
structures at the various scales to fuel nuclear starbursts and AGN.
At scales $\sim$ 1-10pc, when the gas fraction is still high,
violent clump instabilities drive turbulence, 
and consequent viscosity and torques. Dynamical friction,
warps and bending instabilities will drive gas to the nucleus, 
form a thick disk/torus and produce chaotic and intermittent fueling.
When the gas is consumed, the nuclear stellar disk remains such 
that in M31.

Given the various dynamical instabilities of
warping and bending, alignment between small scale and large scale disks
is not expected.
The fueling is certainly moderated by 
feedback processes, from star formation or the AGN itself.
Their efficiency is still poorly known, although gas outflows are present, 
sometimes with huge amounts, up to three times the star formation rate.

%\subsection{Acknowledgments}
%I want to thank the organisers for the invitation to such an interesting conference.

\end{document}